\documentclass[prl,twocolumn]{revtex4-2}
\usepackage[utf8]{inputenc}
\usepackage[T1]{fontenc}
\newcommand{\ud}{\mathrm{d}}
\usepackage{amsmath}
\usepackage{amssymb}
\usepackage{bbm}
\usepackage{newtxtext}
\usepackage[smallerops]{newtxmath}
\let\lambda\lambdaup
\allowdisplaybreaks
\usepackage{graphicx}
\graphicspath{{fig/}}
\usepackage{xcolor}
\usepackage[colorlinks=true,linkcolor=blue,citecolor=blue,urlcolor=blue]{hyperref}
\usepackage[capitalise]{cleveref}
\usepackage{multirow}

\newcommand{\z}[1]{\mathbb{Z}_{#1}}
\newcommand{\unity}{\mathbbm{1}}

\begin{document}

\title{Parafermionic and decoupled multicritical points in a frustrated $\mathbb{Z}_6$ clock chain}
\author{Andrea Kouta Dagnino}
\author{Attila Szabó}
\affiliation{Physik-Institut, Universität Zürich, Winterthurerstr.\ 190, 8057 Zürich, Switzerland}
\date{\today}

\begin{abstract}
We introduce a generalised six-state clock chain that interpolates between the clock and Potts models via a multicritical point described by decoupled Ising and three-state Potts models.
We find that this decoupling extends into stable phases that break only $\z2$ or $\z3$ symmetry.
We also use boundary CFT analysis and level spectroscopy to conclusively identify a $\z6$ parafermion multicritical point terminating the clock model Luttinger-liquid phase.
Our work shows that parafermions emerge far from integrability, even in systems with intertwined Ising and three-state Potts orders.
\end{abstract}

\maketitle


\begin{figure*}
    \centering
    \includegraphics{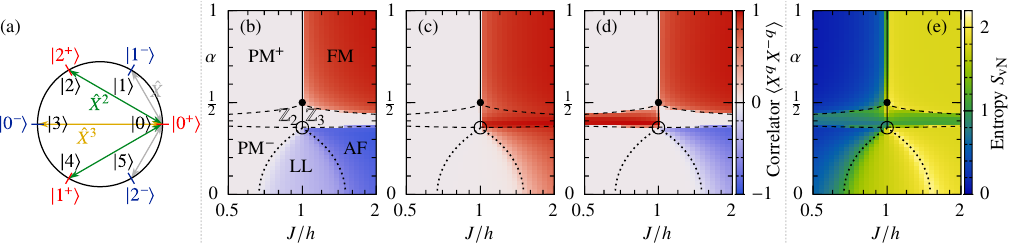}
    \caption{(a) The on-site Hilbert space (in the $\hat Z$ eigenbasis) of the $\z6$ clock model can be decomposed into an Ising ($\pm$, red and blue) and a three-state Potts (numbers $0,1,2$) degree of freedom. At $\alpha=1/2$, the $XX^\dagger$ terms (grey arrows) cancel in~\eqref{eq: combined H}; the remaining $X^2X^{-2}$ (green) and $X^3X^3$ (gold) terms act only on the Potts and Ising degrees of freedom, respectively.
    (b--d) $\langle XX^\dagger\rangle$, $\langle X^2X^{-2}\rangle$, and $\langle X^3X^3\rangle$ correlation functions of~\eqref{eq: combined H} on a 240-site chain between sites 60 and 181. Red and blue indicate ferromagnetic and antiferromagnetic correlations, respectively.
    (e) Half-chain entanglement entropy of the same ground states.
    Solid, dashed, and dotted lines indicate first-order, Ising, and BKT transitions, respectively (the latter two only approximately).
    Phase labels: 
    PM$^\pm$ -- paramagnet dominated by the $|0^\pm\rangle$ state; 
    FM, AF -- $\z6$ broken (anti)ferromagnet;
    $\z{2,3}$ -- partially ordered phase breaking the $\z{2,3}$ subgroup of $\z6$;
    LL -- Luttinger liquid.
    }
    \label{fig: overview}
\end{figure*}

The two-dimensional Ising and three-state Potts models, along with their one-dimensional quantum counterparts, have been standard settings for two-dimensional conformal field theory (CFT) ever since the inception of the field~\cite{Belavin1984CFT,Friedan1984CFT,DotsenkoFateev1984,Dotsenko1984Potts3,FateevZamolodchikov1987Z3,Cardy1986OperatorContent}.
There are several ways to generalise these models to systems with $p>3$-dimensional local Hilbert spaces, two of which are commonly studied. The first is the \textit{$p$-state clock model:} in the quantum chain formulation, its Hamiltonian is
\begin{equation}
    H_\mathrm{clock} = - J \sum_{i=1}^{L-1} \big(\hat X_i^{\dagger} \hat X_{i+1} + \hat X_i \hat X_{i+1}^{\dagger}\big) - h \sum_{i=1}^L \big(\hat Z_i + \hat Z_i^\dagger\big),
    \label{eq: clock}
\end{equation}
where the operators $\hat X_i$ and $\hat Z_i$ obey
\begin{align}
    \hat X_i^p = \hat Z_i^p &= \unity, &
    \hat X_i \hat Z_j &= \begin{cases}
        \omega \hat Z_j \hat X_i & i=j\\
        \hat Z_j \hat X_i & i\neq j.
    \end{cases} &
    (\omega &= e^{2\pi i/p})
    \label{eq: commutation relation}
\end{align}
The second is the \textit{$p$-state Potts model:}
\begin{equation}
    H_\mathrm{Potts} = - J \sum_{i=1}^{L-1} \sum_{q=1}^{p-1} X_i^{-q} X_{i+1}^q - h \sum_{i=1}^L\sum_{q=1}^{p-1} Z_i^q,
    \label{eq: Potts}
\end{equation}
which is symmetric under all global permutations of $\hat X$ eigenstates, not just the $\z p$ subgroup of cyclic permutations.
For $p=4$, both models have a paramagnetic (PM) and a ferromagnetic (FM) phase separated by a conformal critical point at $J=h$~\cite{Baxter1971Eight-vertexStatistics,KadanoffWagner1971,Fan1972AshkinTeller, KadanoffBrown1979,KohmotoDenNijsKadanoff1981}.
For $p\ge5$, however, the two models behave qualitatively differently:
The Potts model undergoes a first-order transition at $J=h$~\cite{Baxter1973PottsCriticalTemperature,Wu1982Potts},
while the clock model exhibits an extended Luttinger-liquid (LL) phase around $J=h$, which is separated from the PM and FM phases by BKT transitions~\cite{Jose1977PlanarModel,Elitzur1979DiscreteAbelian,Ortiz2012PClockDualities, Sun2019ZpClockTransitions}.

The clock model can naturally be generalised by adding all other nearest-neighbour couplings that respect its $D_{2p}$ symmetry:
\begin{equation}
    H = -\sum_{q=1}^{p-1} \bigg[ J_q \sum_{i=1}^{L-1} X_i^{-q} X_{i+1}^q + h_q \sum_{i=1}^L Z_i^q \bigg],
    \label{eq: most general H}
\end{equation}
with real $J_q=J_{p-q}$ and $h_q=h_{p-q}$ for achiral models.
For $p=4$, this Hamiltonian generates the Ashkin--Teller line of critical points with continuously varying exponents~\cite{Baxter1971Eight-vertexStatistics,KadanoffWagner1971,Fan1972AshkinTeller, KadanoffBrown1979,KohmotoDenNijsKadanoff1981}.
For $p>4$, only the integrable point $J_q=h_q\propto 1/\sin(\pi q/p)$~\cite{Fateev1982IntegrableZN,Alcaraz1986QuantumIntegrableZN,Zamolodchikov1985Parafermion,JimboMiwaOkado1986BrokenZN,Albertini1994FZSpinChain}, described by the $\z p$ parafermion CFT~\cite{Zamolodchikov1985Parafermion,Fradkin1980Parafermion}, has been studied in detail.
For $p=5$, this is a multicritical point between the BKT and first-order transitions of~\mbox{(\ref{eq: clock},\,\ref{eq: Potts})}, essentially fixing the phase diagram~\cite{Zamolodchikov1985Parafermion}.

We therefore focus on the next smallest case, $p=6$ \cite{Challa1986SixStateClock,Matsuo2006SixStateClock},
which is also remarkable in that $D_{12} = \z2\times S_3$, which suggests interpreting the on-site Hilbert space as an Ising and a three-state Potts degree of freedom.
We make this interpretation explicit through the Hamiltonian
\begin{equation}
    H_\alpha = \alpha H_\mathrm{Potts} - (1-\alpha) H_\mathrm{clock},
    \label{eq: combined H}
\end{equation}
which interpolates between the Potts model ($\alpha=1$) and the \textit{antiferromagnetic} clock model ($\alpha=0$; equivalent to the ferromagnetic clock model, see End Matter).
Notably, at $\alpha=1/2$, the $q=1,5$ terms of~\eqref{eq: most general H} cancel out: As shown in \cref{fig: overview}(a), the remaining terms form decoupled Ising ($q=3$) and three-state Potts ($q=2,4$) chains, so the model undergoes a direct second-order $\z6$-breaking transition in the $\mathrm{Ising\times Potts_3}$ universality class.
Together with the first-order transition at $\alpha=1$ and the intermediate LL phase at $\alpha=0$, this already hints at a complex phase diagram as a function of $\alpha$.

In what follows, we uncover an even richer phase diagram using large-scale DMRG simulations. 
We find that near $\alpha=1/2$, the $\z2$ and $\z3$ subgroups of the global $\z6$ symmetry decouple, with phases that break only one or the other emerging from the multicritical point $\alpha=1/2,J=h$.
We also conclusively place the multicritical point that terminates the LL phase of the clock model in the $\z6$ parafermion universality class:
Given the adjacent LL phases and BKT transitions with qualitatively different critical scaling, this requires novel, sophisticated analysis of DMRG results with both open and periodic boundaries.
The appearance of a parafermionic critical point very far from the integrable Fateev--Zamolodchikov point is a remarkable deviation from the usual expectation~\cite{Zamolodchikov1985Parafermion}:
We end the paper with a brief discussion of further, experimentally more easily realisable scenarios where $\z6$ parafermions may emerge.


\textit{$\z2\times\z3$ multicriticality at $\alpha=1/2$.---}
We start exploring the phase diagram by perturbing the $\mathrm{Ising\times Potts_3}$ multicritical point at $\alpha=1/2, J=h$.
There are three relevant ($\Delta<2$) operators in the theory that can appear as additional terms in the Hamiltonian (i.e., preserve the full $\mathbb{Z}_2 \times S_3$ symmetry group of the model and have zero conformal spin): The thermal operators of the Ising and $\mathrm{Potts_3}$ components, $\Psi_1=\varepsilon_I$ and $\Psi_2=\varepsilon_P$, and their product, $\Psi_3 = \varepsilon_I\otimes \varepsilon_P$ (scaling dimensions: 1, 0.8, 1.8).
Consider now the effective action
\begin{equation}
S = S_\text{CFT} + \sum_{i=1}^3 g_i \int \ud^2x\, \Phi_i(x),
\end{equation}
where $S_\text{CFT}$ is the action of the unperturbed $\mathrm{Ising\times Potts_3}$ CFT.
The renormalisation-group flow of the coupling constants $g_i$ follows from standard conformal perturbation theory~\cite{Zamolodchikov1986CTheorem,Cardy1988LesHouches} as
\begin{equation}
    \begin{cases}
        \frac{\ud g_1}{\ud\ell} = g_1 - 2\pi g_2 g_3 + O(g^3)\\
         \frac{\ud g_2}{\ud\ell} =\frac{6}{5} g_2 - 2\pi g_1 g_3 + O(g^3)\\
         \frac{\ud g_3}{\ud\ell} = \frac{1}{5} g_3 - 2\pi g_1 g_2 + O(g^3).
    \end{cases}
    \label{eq: RG flow}
\end{equation}
This flow has no stable fixed point; for almost all initial parameters, it diverges to large $g_i$ that satisfy $g_1g_2g_3<0$.

In the phase diagram cut of \cref{fig: overview}, the Ising and Potts components are perturbed away from criticality by the same amount, i.e., we set $g_1=g_2$ initially.
Now, if $g_3 < 0$, $g_{1,2}$ diverge without changing sign, that is, the RG flows to either a fully symmetric paramagnet ($g_{1,2}\to-\infty$) or a $\z6$-broken ferromagnet ($g_{1,2}\to+\infty$): this is consistent with $\alpha>1/2$.
By contrast, if $g_3>0$, $g_1$ and $g_2$ must have opposite signs at late RG times, i.e., the system breaks only $\z2$ or $\z3$ symmetry, as seen at $\alpha<1/2$.
The initial equality of $g_{1,2}$ is lifted by the linear terms in~\eqref{eq: RG flow}: 
If $g_{1,2}>0$, the higher coefficient of $g_2$ will cause it to grow more positive, which ultimately leads to $g_2\to+\infty,g_1\to-\infty$, i.e., a $\z3$ broken state.
Likewise, if $g_{1,2}<0$, $g_2$ grows more negative, resulting in $g_2\to-\infty,g_1\to+\infty$, i.e., a $\z2$ broken state,
consistent with the layout of the phase diagram in \cref{fig: overview}.
We anticipate a second-order Ising transition between the paramagnet and $\z2$, as well as between the $\z3$ and $\z6$, phases, and first-order transitions along $J=h$: this is indeed borne out by the order parameters in \cref{fig: overview}.

\begin{figure}
    \centering
    \includegraphics{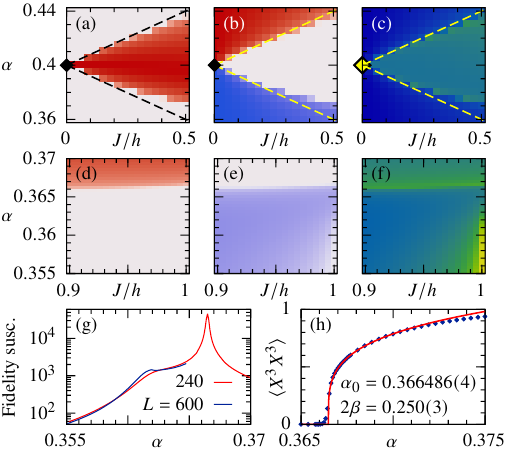}
    \caption{Ising correlator $\langle X^3_{60}X^3_{181}\rangle$ (a), dual Ising string correlator $\langle \prod_{i=60}^{180} Z_i^3\rangle$ (b), and half-chain entanglement entropy (c) of~\eqref{eq: combined H} on a 240-site chain at low $J/h$ near the spin-flop transition at $J=0,\alpha=2/5$ (diamond).
    Dashed lines indicate the critical lines of the effective transverse-field Ising model~\eqref{eq: effective Ising}.
    (d--f) The same quantities in the confluence region $J\approx h, \alpha\approx 0.36$ (empty circle in \cref{fig: overview}). 
    Colour maps match those in \cref{fig: overview}.
    (g) Fidelity susceptibility along the $J=h$ line for 240-site and 600-site chains, indicating two transitions at $\alpha\approx0.361,0.3665$.
    (h) Ising correlator $\langle X^3_{60}X^3_{181}\rangle$ at $J=h$ (blue dots) with a power-law fit (red).
    }
    \label{fig: Ising lobe}
\end{figure}

We next consider the fate of the $\z2$-broken phase at low $J/h$.
(By Kramers--Wannier duality, this is also equivalent to the $\z3$-broken phase at high $J/h$.)
At $J=0$, lattice sites decouple completely, so the ground state is a $Z$-basis product state, selected by the single-site terms:
The optimal state flops from $|0\rangle\equiv |0^+\rangle$ to $|3\rangle\equiv |0^-\rangle$ at $\alpha_0=2/5$.
Turning on a finite $J>0$ couples these states: To lowest order in perturbation theory, the system is described by the transverse-field Ising chain
\begin{equation}
    H_0 = \alpha_0 J\sum_i \sigma_i^x \sigma_{i+1}^x - 5(\alpha-\alpha_0)h \sum_i \sigma_i^z,
    \label{eq: effective Ising}
\end{equation}
where $|\sigma^z=\pm1\rangle\equiv |0^{\pm}\rangle$.
This model has Ising transitions at $J/h = 12.5|\alpha-\alpha_0|$, which matches the numerically obtained phase boundaries in \cref{fig: Ising lobe}(a--c).
As shown in \cref{fig: Ising lobe}(h), the Ising nature of the lower transition line then persists all the way until it meets the first-order line at $J=h$ around $\alpha=0.3665$.


\textit{Luttinger liquids (LL) and parafermionic multicriticality.---}
Let us now approach the confluence region around $J=h, \alpha=0.36$ (empty circle in \cref{fig: overview}) from the side of small $\alpha$. 
In the clock model~\eqref{eq: clock} at $\alpha=0$, there is an extended LL phase around $J=h$, where both regular and string correlators decay algebraically (cf.~\cref{fig: overview}), bounded by BKT transitions on either side.
This phase structure persists for finite $\alpha$. 
However, the LL phase becomes narrower and closes at a multicritical point $\alpha\approx0.361$ [\cref{fig: Ising lobe}(d--g)]:
Here, both BKT transitions meet the $J=h$ first-order line, but not the Ising lines that bound the $\z{2,3}$ phases.
In the following, we will determine the universality class of this multicritical point.


\begin{figure}
    \centering
    \includegraphics{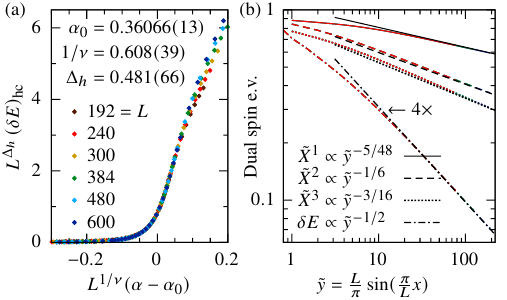}
    \caption{DMRG results with open boundaries near the parafermionic multicritical point.
    (a) Finite-size scaling of the deviation $\delta E = \langle h^{(2)}\rangle -\langle h^{(1)}\rangle$ from self-duality for open chains at $J=h$.
    (b) $\delta E$ and dual spin expectation values as a function of the distance $x$ from the end of the chain for three values of $L$ at $\alpha=0.3606$, $J=h$, with fits to the relevant $\z6$ parafermion CFT scaling dimensions. 
    }
    \label{fig: OBC}
\end{figure}

\textit{Open chains.---}
We first performed DMRG simulations with open boundary conditions along the self-dual line $J=h$ between $0.35\le\alpha\le 0.366$ for six system sizes $192\le L\le 600$.  
We found that the expectation values of the one-body terms $\langle h^{(1)}\rangle$ of the Hamiltonian are consistently more negative than those of the two-body terms, $\langle h^{(2)}\rangle$.
While this indicates duality breaking at all $\alpha$, the discrepancy is smaller and tends to zero with increasing $L$ for $\alpha\lesssim0.36$, but remains finite along the first-order phase boundary.
Therefore, we can use the deviation at the middle of the chain, $\delta E:=\langle h^{(2)}_{L/2,L/2+1}\rangle - \langle h^{(1)}_{L/2}\rangle$, as an order parameter of the LL--PM transition at $J=h^-$.
As shown in \cref{fig: OBC}(a), $\delta E$ obeys a finite-size scaling law, which pinpoints the multicritical point to $\alpha_0=0.36066(13)$ (rounded to $0.3606$ below). However, the critical exponents have far too large error bars to allow characterising the critical theory.

Usually, one would obtain further critical exponents from the scaling of correlation functions such as $\langle X^q_i X^{-q}_j\rangle$. 
Here, however, strong boundary effects and slow convergence to the asymptotic power law (indicating close subleading scaling dimensions) prevent us from obtaining accurate critical exponents even at $L=600$.
Instead, we focus on one-point functions of the underlying field theory: While these must vanish at the multicritical point in periodic boundary conditions,
open boundaries that break (generalised) symmetries may allow them to develop an expectation value with spatial dependence
\begin{equation}
    \langle \phi(x)\rangle \propto \tilde y^{-\Delta_\phi};\quad
    \tilde y := \frac{L}\pi \sin\left(\frac{\pi x}L \right),
\end{equation}
where $\Delta_\phi$ is the (bulk) scaling dimension of $\phi$~\cite{Francesco1997YellowBook}.

In our spin chains, introducing open boundary conditions at $J=h$ breaks both the dual $\z6$ symmetry and the duality $J\leftrightarrow h$ in favour of a trivial paramagnet, where the dual order parameters $\tilde X^q$~\eqref{eq: dual OP} develop finite expectation values.
\cref{fig: OBC}(b) shows that these are $L$-independent universal functions of the renormalised distance $\tilde y$ from the boundary;
fitting power laws to them at large $\tilde y$ yields exponents close to the order-parameter scaling dimensions $5/48,1/6,3/32$ of the $\z6$ parafermion CFT~\cite{Zamolodchikov1985Parafermion}.
Likewise, the expectation value of the duality-breaking primary operator can be estimated from the deviations $\delta E(i-\frac12) = \frac{E}{2L} - \langle h^{(1)}_i\rangle$, $\delta E(i) = \langle h^{(2)}_{i,i+1}\rangle-\frac{E}{2L}$ of Hamiltonian terms from their bulk value: These also scale as a power law of $\tilde y$ with exponent close to $1/2$, the lowest thermal scaling dimension of the parafermion CFT.


\begin{figure}
    \centering
    \includegraphics{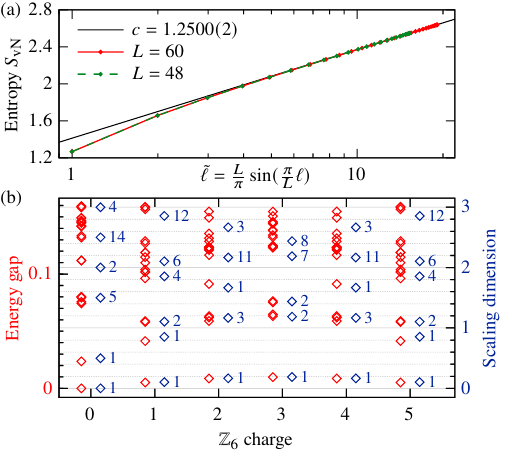}
    \caption{DMRG results with periodic boundary conditions at the parafermionic multicritical point $\alpha=0.3606, J=h$.
    (a) Entanglement entropy $S_\mathrm{vN}$ as a function of subsystem size $\ell$, with a fit to the CFT prediction $S_\mathrm{vN} = c\log (\tilde\ell)/3+\mathrm{const.}$
    (b) Lanczos spectrum of the mid-chain DMRG step for $L=60$ (red), compared to the scaling dimensions and multiplicities of the $\z6$ parafermion CFT (blue).
    }
    \label{fig: PBC}
\end{figure}

\textit{Periodic boundary conditions.---}
Nevertheless, the duality-breaking open boundaries, combined with the qualitatively different critical behaviour of the adjacent LL phases and BKT transition lines, make it impossible to accurately extract additional signatures of the conformal multicritical point, such as the central charge, from simulations on open chains.
Therefore, we also performed DMRG calculations for rings of length $L=48,60$ with periodic boundary conditions at the multicritical point $\alpha=0.3606, J=h$, using bond dimension $\chi=1920$ to account for the higher entanglement.
As expected, the converged ground states are self-dual to a good approximation; their entanglement entropy follows the CFT scaling law $S=\frac{c}3\log\tilde \ell+\mathrm{const.}$~\cite{CalabreseCardy2004PBC} [\cref{fig: PBC}(a)] with the $\z6$ parafermion  central charge $c=1.25$.

Finally, following~\cite{Chepiga2017SpectrumDMRG}, we estimated the low-energy excitation spectrum of the Hamiltonian using the spectrum of the effective Hamiltonian constructed by a DMRG step at the midpoint of the chain (see End Matter for details).
This effectively truncates the Hilbert space of the system to that spanned by the leading Schmidt vectors of the ground state;
at a conformal critical point, however, low-energy excited states have a high overlap with this subspace~\cite{Chepiga2017SpectrumDMRG}.
As shown in \cref{fig: PBC}(b), the layout and approximate multiplicities of the spectrum match that of the bulk $\z6$ parafermion CFT quite accurately up to $\Delta\lesssim2$. 
Altogether, our DMRG simulations with both open and periodic boundary conditions provide ample evidence for a multicritical point at $\alpha\approx0.3606,J=h$ described by the $\z6$ parafermion CFT.


\textit{Discussion.---}
In summary, we studied the remarkably rich ground-state phase diagram of a generalised six-state clock chain that interpolates between the clock and Potts models, as well as a multicritical point described by an $\mathrm{Ising\times Potts_3}$ CFT.
The decoupling of the $\z2$ and $\z3$ subgroups of the overall $\z6$ symmetry group defines the phase diagram far away from the critical point, opening up lobes of reentrant $\z2$ and $\z3$ symmetry breaking.

In addition, we find a Luttinger-liquid phase stretching from the clock model up to a multicritical point:
Combining excited-state spectra with a novel analysis of open-boundary DMRG data that exploits, rather than minimises the effect of, symmetry-breaking boundary conditions, we conclusively identify $\z6$ parafermionic CFT at this multicritical point, for the first time on a spin chain to our knowledge. 
(Anyon chains \textit{designed} to exhibit parafermions have, however, been studied in~\cite{gils2013anyonic}.)
Remarkably, we did so far away from known integrable points~\cite{Fateev1982IntegrableZN,Alcaraz1986QuantumIntegrableZN,Zamolodchikov1985Parafermion,JimboMiwaOkado1986BrokenZN,Albertini1994FZSpinChain},  where parafermionic criticality was widely believed to be restricted.

Higher-order parafermionic criticality thus appears much more common than previously assumed.
In particular, $\z6$ parafermions probably emerge along a multicritical \textit{line} in the phase diagram of~\eqref{eq: most general H}, which includes the multicritical point of the ferromagnetic clock--Potts interpolation Hamiltonian~\cite{Polackova2023CTMRG}.
More interestingly, our results highlight that intertwined Ising and three-state Potts order parameters can readily combine into effective six-state clock variables [\cref{fig: overview}(a)] and thus exhibit $\z6$ parafermions and LL phases, which are not expected from either order parameter individually.
A case in point is the frustrated triangular-lattice Ising model studied in~\cite{Rakala2021TriangularIsingZ6}, where a combination of Ising-symmetry breaking and three-sublattice spatial order on the triangular lattice results in $\z6$ parafermionic criticality.

In a similar fashion, it may prove easier to realise $\z6$ parafermionic spin chains out of interacting qubit and qutrit degress of freedom \cite{Fendley2012ParafermionicEdge,AliceaFendley2016Parafermions,Goss2022QutritGates,Iqbal2025QutritToric,
Ringbauer2022QuditProcessor}.
Nevertheless, six-dimensional local Hilbert spaces are readily available in the hyperfine levels of such atoms as $^{173}$Yb~\cite{Ringbauer2022QuditProcessor,CazalillaRey2014SUN,
Taie2010SU2SU6,Taie2012SU6Mott,Pagano2014OneDimLiquid,
Scazza2014TwoOrbitalSpinExchange,Taie2022AntiferromagneticSUN}, potentially allowing the necessary Hamiltonian terms to be engineered in atom trap setups.
It may also be possible to realise the equivalent statistical mechanics models in artificial spin-ice lattices~\cite{Skjaervo2020ArtificialSpinIce} with sixfold symmetric nanomagnets.

In future work, we plan to explore the broader phase diagram of the generalised clock model~\eqref{eq: most general H}.
In particular, the close approach of the LL and partially symmetry-broken phases suggests that they touch at a nearby higher-order multicritical point. 
This may lead to further exotic phases and transitions, especially as the frustration between the opposite-sign $J_1$ and $J_3$ terms comes to dominate the Hamiltonian.
It will also be interesting to deploy coupled-wire constructions on these spin chains to build systems with parafermionic \textit{topological order,} a promising building block of topological quantum computation~\cite{Hutter2016QuantumComputingParafermion}.

\begin{acknowledgements}
    We thank Zheng-Cheng Gu and Titus Neupert for helpful discussions.
    DMRG simulations were performed using the TeNPy~\cite{tenpy} library.
    The finite-size scaling analysis in \cref{fig: OBC}(a) was done using the pyfssa~\cite{pyfssa} library.
    All heat maps use perceptionally uniform colour maps developed in Ref.~\cite{colorcet}.
    A.\,Sz. was supported by Ambizione grant No. 215979 by the Swiss National Science Foundation.
    A.\,K.\,D. acknowledges support from the Swiss National Science Foundation through a Consolidator Grant (iTQC, TMCG-2213805).
\end{acknowledgements}

\bibliography{attila_everything,bib,parafermion,cft,software,kouta}

@article{Baxter1971Eight-vertexStatistics,
    title = {{Eight-vertex model in lattice statistics}},
    year = {1971},
    journal = {Phys. Rev. Lett.},
    author = {Baxter, R. J.},
    number = {14},
    month = {4},
    pages = {832--833},
    volume = {26},
    publisher = {American Physical Society},
    url = {https://journals.aps.org/prl/abstract/10.1103/PhysRevLett.26.832},
    doi = {10.1103/PhysRevLett.26.832},
    issn = {00319007}
}

@Article{Skjaervo2020ArtificialSpinIce,
author={Skj{\ae}rv{\o}, Sandra H.
and Marrows, Christopher H.
and Stamps, Robert L.
and Heyderman, Laura J.},
title={Advances in artificial spin ice},
journal={Nat. Rev. Phys.},
year={2020},
month={Jan},
day={01},
volume={2},
number={1},
pages={13-28},
issn={2522-5820},
doi={10.1038/s42254-019-0118-3},
url={https://doi.org/10.1038/s42254-019-0118-3}
}

@book{Francesco1997YellowBook,
title="Conformal field theory",
author="Philippe Francesco and Pierre Mathieu and David Sénéchal",
doi="https://doi.org/10.1007/978-1-4612-2256-9",
series="Graduate Texts in Contemporary Physics",
year=1997,
publisher="Springer-Verlag",
city="New York",
isbn="978-1-4612-2256-9"
}

@article{Hutter2016QuantumComputingParafermion,
  title = {Quantum computing with parafermions},
  author = {Hutter, Adrian and Loss, Daniel},
  journal = {Phys. Rev. B},
  volume = {93},
  issue = {12},
  pages = {125105},
  numpages = {7},
  year = {2016},
  month = {Mar},
  publisher = {American Physical Society},
  doi = {10.1103/PhysRevB.93.125105},
  url = {https://link.aps.org/doi/10.1103/PhysRevB.93.125105}
}

@article{Singh2011U1SymmetricMPS,
  title = {Tensor network states and algorithms in the presence of a global {U(1)} symmetry},
  author = {Singh, Sukhwinder and Pfeifer, Robert N. C. and Vidal, Guifre},
  journal = {Phys. Rev. B},
  volume = {83},
  issue = {11},
  pages = {115125},
  numpages = {22},
  year = {2011},
  month = {Mar},
  publisher = {American Physical Society},
  doi = {10.1103/PhysRevB.83.115125},
  url = {https://link.aps.org/doi/10.1103/PhysRevB.83.115125}
}

@article{Belavin1984CFT,
  author  = {Belavin, A. A. and Polyakov, A. M. and Zamolodchikov, A. B.},
  title   = {Infinite conformal symmetry in two-dimensional quantum field theory},
  journal = {Nucl. Phys. B},
  volume  = {241},
  number  = {2},
  pages   = {333--380},
  year    = {1984},
  doi     = {10.1016/0550-3213(84)90052-X}
}

@article{Friedan1984CFT,
  title = {Conformal Invariance, Unitarity, and Critical Exponents in Two Dimensions},
  author = {Friedan, Daniel and Qiu, Zongan and Shenker, Stephen},
  journal = {Phys. Rev. Lett.},
  volume = {52},
  issue = {18},
  pages = {1575--1578},
  numpages = {0},
  year = {1984},
  month = {Apr},
  publisher = {American Physical Society},
  doi = {10.1103/PhysRevLett.52.1575},
  url = {https://link.aps.org/doi/10.1103/PhysRevLett.52.1575}
}

@article{Cardy1986OperatorContent,
    author = "Cardy, John L.",
    title = "Operator Content of Two-Dimensional Conformally Invariant Theories",
    doi = "10.1016/0550-3213(86)90552-3",
    journal = "Nucl. Phys. B",
    volume = "270",
    pages = "186--204",
    year = "1986"
}

@article{CalabreseCardy2004PBC,
doi = {10.1088/1742-5468/2004/06/P06002},
url = {https://doi.org/10.1088/1742-5468/2004/06/P06002},
year = {2004},
month = {jun},
publisher = {},
volume = {2004},
number = {06},
pages = {P06002},
author = {Pasquale Calabrese and John Cardy},
title = {Entanglement entropy and quantum field theory},
journal = {J. Stat. Mech.}
}

@article{Dotsenko1984Potts3,
    author = "Dotsenko, V. S.",
    title = "{Critical Behavior and Associated Conformal Algebra of the Z(3) {P}otts Model}",
    doi = "10.1016/0550-3213(84)90148-2",
    journal = "Nucl. Phys. B",
    volume = "235",
    pages = "54--74",
    year = "1984"
}

@article{Chepiga2017SpectrumDMRG,
  title = {Excitation spectrum and density matrix renormalization group iterations},
  author = {Chepiga, Natalia and Mila, Fr\'ed\'eric},
  journal = {Phys. Rev. B},
  volume = {96},
  issue = {5},
  pages = {054425},
  numpages = {16},
  year = {2017},
  month = {Aug},
  publisher = {American Physical Society},
  doi = {10.1103/PhysRevB.96.054425},
  url = {https://link.aps.org/doi/10.1103/PhysRevB.96.054425}
}

@ARTICLE{Zamolodchikov1986CTheorem,
       author = {{Zamolodchikov}, A. B.},
        title = "{``Irreversibility''} of the flux of the renormalization group in a {2D} field theory",
      journal = {JETP Lett.},
         year = 1986,
        month = jun,
       volume = {43},
        pages = {730},
       adsurl = {https://ui.adsabs.harvard.edu/abs/1986JETPL..43..730Z},
      url = {http://www.jetpletters.ru/ps/1413/article_21504.pdf}
}

@inproceedings{Cardy1988LesHouches,
    author = "John L. Cardy",
    title = "Conformal invariance and statistical mechanics",
    booktitle = {Les Houches Summer School in Theoretical Physics: Fields, Strings, Critical Phenomena},
    year = 1988,
    volume = 49,
    url = "https://www-thphys.physics.ox.ac.uk/people/JohnCardy/lh.pdf"
}

@article{DotsenkoFateev1984,
  author  = {Dotsenko, Vl. S. and Fateev, V. A.},
  title   = {Conformal algebra and multipoint correlation functions in {2D} statistical models},
  journal = {Nucl. Phys. B},
  volume  = {240},
  number  = {3},
  pages   = {312--348},
  year    = {1984},
  doi     = {10.1016/0550-3213(84)90269-4}
}

@article{Wu1982Potts,
  author  = {Wu, F. Y.},
  title   = {The {Potts} model},
  journal = {Rev. Mod. Phys.},
  volume  = {54},
  number  = {1},
  pages   = {235--268},
  year    = {1982},
  doi     = {10.1103/RevModPhys.54.235}
}

@article{FateevZamolodchikov1987Z3,
  author  = {Fateev, V. A. and Zamolodchikov, A. B.},
  title   = {Conformal quantum field theory models in two dimensions having $\mathbb{Z}_3$ symmetry},
  journal = {Nucl. Phys. B},
  volume  = {280},
  pages   = {644--660},
  year    = {1987},
  doi     = {10.1016/0550-3213(87)90183-0}
}

@article{Baxter1973PottsCriticalTemperature,
  author  = {Baxter, R. J.},
  title   = {Potts model at the critical temperature},
  journal = {J. Phys. C},
  volume  = {6},
  number  = {23},
  pages   = {L445--L448},
  year    = {1973},
  doi     = {10.1088/0022-3719/6/23/005}
}

@article{Elitzur1979DiscreteAbelian,
  title = {Phase structure of discrete Abelian spin and gauge systems},
  author = {Elitzur, S. and Pearson, R. B. and Shigemitsu, J.},
  journal = {Phys. Rev. D},
  volume = {19},
  issue = {12},
  pages = {3698--3714},
  numpages = {0},
  year = {1979},
  month = {Jun},
  publisher = {American Physical Society},
  doi = {10.1103/PhysRevD.19.3698},
  url = {https://link.aps.org/doi/10.1103/PhysRevD.19.3698}
}

@article{KohmotoDenNijsKadanoff1981,
  title = {Hamiltonian studies of the $d=2$ Ashkin-Teller model},
  author = {Kohmoto, Mahito and den Nijs, Marcel and Kadanoff, Leo P.},
  journal = {Phys. Rev. B},
  volume = {24},
  issue = {9},
  pages = {5229--5241},
  numpages = {0},
  year = {1981},
  month = {Nov},
  publisher = {American Physical Society},
  doi = {10.1103/PhysRevB.24.5229},
  url = {https://link.aps.org/doi/10.1103/PhysRevB.24.5229}
}

@article{Fan1972AshkinTeller,
  author  = {Fan, Chungpeng},
  title   = {On critical properties of the {Ashkin--Teller} model},
  journal = {Phys. Lett. A},
  volume  = {39},
  number  = {2},
  pages   = {136},
  year    = {1972},
  doi     = {10.1016/0375-9601(72)91051-1}
}

@article{KadanoffBrown1979,
  author  = {Kadanoff, Leo P. and Brown, Alan C.},
  title   = {Correlation functions on the critical lines of the {Baxter} and {Ashkin--Teller} models},
  journal = {Ann. Phys.},
  volume  = {121},
  number  = {2},
  pages   = {318--342},
  year    = {1979},
  doi     = {10.1016/0003-4916(79)90100-3}
}

@article{KadanoffWagner1971,
  title = {Some Critical Properties of the Eight-Vertex Model},
  author = {Kadanoff, Leo P. and Wegner, Franz J.},
  journal = {Phys. Rev. B},
  volume = {4},
  issue = {11},
  pages = {3989--3993},
  numpages = {0},
  year = {1971},
  month = {Dec},
  publisher = {American Physical Society},
  doi = {10.1103/PhysRevB.4.3989},
  url = {https://link.aps.org/doi/10.1103/PhysRevB.4.3989}
}

@article{Fendley2012ParafermionicEdge,
doi = {10.1088/1742-5468/2012/11/P11020},
url = {https://doi.org/10.1088/1742-5468/2012/11/P11020},
year = {2012},
month = {nov},
publisher = {IOP Publishing and SISSA},
volume = {2012},
number = {11},
pages = {P11020},
author = {Fendley, Paul},
title = {Parafermionic edge zero modes in {$Z_n$}-invariant spin chains},
journal = {J. Stat. Mech.}
}

@article{AliceaFendley2016Parafermions,
   author = "Alicea, Jason and Fendley, Paul",
   title = "Topological Phases with Parafermions: Theory and Blueprints", 
   journal= "Annu. Rev. Condens. Matter Phys.",
   year = "2016",
   volume = "7",
   number = "Volume 7, 2016",
   pages = "119-139",
   doi = "https://doi.org/10.1146/annurev-conmatphys-031115-011336",
   url = "https://www.annualreviews.org/content/journals/10.1146/annurev-conmatphys-031115-011336",
   publisher = "Annual Reviews",
   issn = "1947-5462",
   type = "Journal Article",
  }

@article{Ringbauer2022QuditProcessor,
  author        = {Ringbauer, Martin and Meth, Michael and Postler, Lukas and Stricker, Roman and Blatt, Rainer and Schindler, Philipp and Monz, Thomas},
  title         = {A universal qudit quantum processor with trapped ions},
  journal       = {Nat. Phys.},
  volume        = {18},
  pages         = {1053--1057},
  year          = {2022},
  doi           = {10.1038/s41567-022-01658-0},
  archivePrefix = {arXiv},
  eprint        = {2109.06903},
  primaryClass  = {quant-ph}
}

@article{Goss2022QutritGates,
  author  = {Goss, Noah and Morvan, Alexis and Marinelli, Brian and Mitchell, Bradley K. and Nguyen, Long B. and Naik, Ravi K. and Chen, Larry and Kreikebaum, John Mark and Santiago, David I. and Wallman, Joel J. and Siddiqi, Irfan},
  title   = {High-fidelity qutrit entangling gates for superconducting circuits},
  journal = {Nat. Commun.},
  volume  = {13},
  pages   = {7481},
  year    = {2022},
  doi     = {10.1038/s41467-022-34851-z}
}

@article{Iqbal2025QutritToric,
  author        = {Iqbal, Mohsin and Lyons, Anasuya and Lo, Chiu Fan Bowen and Tantivasadakarn, Nathanan and Dreiling, Joan and Foltz, Cameron and Gatterman, Thomas M. and Gresh, Dan and Hewitt, Nathan and Holliman, Craig A. and Johansen, Jacob and Neyenhuis, Brian and Matsuoka, Yohei and Mills, Michael and Moses, Steven A. and Siegfried, Peter and Vishwanath, Ashvin and Verresen, Ruben and Dreyer, Henrik},
  title         = {Qutrit toric code and parafermions in trapped ions},
  journal       = {Nat. Commun.},
  volume        = {16},
  pages         = {6301},
  year          = {2025},
  doi           = {10.1038/s41467-025-61391-z},
  archivePrefix = {arXiv},
  eprint        = {2411.04185},
  primaryClass  = {quant-ph}
}

@article{CazalillaRey2014SUN,
  author        = {Cazalilla, Miguel A. and Rey, Ana Maria},
  title         = {Ultracold {Fermi} gases with emergent {$SU(N)$} symmetry},
  journal       = {Rep. Prog. Phys.},
  volume        = {77},
  number        = {12},
  pages         = {124401},
  year          = {2014},
  doi           = {10.1088/0034-4885/77/12/124401},
  archivePrefix = {arXiv},
  eprint        = {1403.2792},
  primaryClass  = {cond-mat.quant-gas}
}

@article{Taie2010SU2SU6,
  title = {Realization of a {$\mathrm{SU}(2)\times\mathrm{SU}(6)$} System of Fermions in a Cold Atomic Gas},
  author = {Taie, Shintaro and Takasu, Yosuke and Sugawa, Seiji and Yamazaki, Rekishu and Tsujimoto, Takuya and Murakami, Ryo and Takahashi, Yoshiro},
  journal = {Phys. Rev. Lett.},
  volume = {105},
  issue = {19},
  pages = {190401},
  numpages = {4},
  year = {2010},
  month = {Nov},
  publisher = {American Physical Society},
  doi = {10.1103/PhysRevLett.105.190401},
  url = {https://link.aps.org/doi/10.1103/PhysRevLett.105.190401},
  archivePrefix = {arXiv},
  eprint        = {1005.3670},
  primaryClass  = {cond-mat.quant-gas}
}

@article{Taie2012SU6Mott,
  author  = {Taie, Shintaro and Yamazaki, Rekishu and Sugawa, Seiji and Takahashi, Yoshiro},
  title   = {An {$SU(6)$} {Mott} insulator of an atomic {Fermi} gas realized by large-spin {Pomeranchuk} cooling},
  journal = {Nat. Phys.},
  volume  = {8},
  pages   = {825--830},
  year    = {2012},
  doi     = {10.1038/nphys2430}
}

@article{Pagano2014OneDimLiquid,
  author  = {Pagano, Guido and Mancini, Marco and Cappellini, Giacomo and Lombardi, Pietro and Sch{\"a}fer, Florian and Hu, Hui and Liu, Xia-Ji and Catani, Jacopo and Sias, Carlo and Inguscio, Massimo and Fallani, Leonardo},
  title   = {A one-dimensional liquid of fermions with tunable spin},
  journal = {Nat. Phys.},
  volume  = {10},
  pages   = {198--201},
  year    = {2014},
  doi     = {10.1038/nphys2878}
}

@article{Scazza2014TwoOrbitalSpinExchange,
  author  = {Scazza, Francesco and Hofrichter, Christian and H{\"o}fer, Moritz and De Groot, Peter C. and Bloch, Immanuel and F{\"o}lling, Simon},
  title   = {Observation of two-orbital spin-exchange interactions with ultracold {$SU(N)$}-symmetric fermions},
  journal = {Nat. Phys.},
  volume  = {10},
  pages   = {779--784},
  year    = {2014},
  doi     = {10.1038/nphys3061}
}

@article{Taie2022AntiferromagneticSUN,
  author        = {Taie, Shintaro and Ibarra-Garc{\'i}a-Padilla, Eduardo and Nishizawa, Naoki and Takasu, Yosuke and Kuno, Yoshihito and Wei, Hao-Tian and Scalettar, Richard T. and Hazzard, Kaden R. A. and Takahashi, Yoshiro},
  title         = {Observation of antiferromagnetic correlations in an ultracold {$SU(N)$} {Hubbard} model},
  journal       = {Nat. Phys.},
  volume        = {18},
  pages         = {1356--1361},
  year          = {2022},
  doi           = {10.1038/s41567-022-01725-6},
  archivePrefix = {arXiv},
  eprint        = {2010.07730},
  primaryClass  = {cond-mat.quant-gas}
}

@article{JimboMiwaOkado1986BrokenZN,
  author  = {Jimbo, M. and Miwa, T. and Okado, M.},
  title   = {Solvable lattice models with broken {$Z_N$} symmetry and {Hecke}'s indefinite modular forms},
  journal = {Nuclear Physics B},
  volume  = {275},
  number  = {3},
  pages   = {517--545},
  year    = {1986},
  doi     = {10.1016/0550-3213(86)90611-5}
}

@article{Albertini1994FZSpinChain,
  author        = {Albertini, Giuseppe},
  title         = {{Fateev--Zamolodchikov} spin chain: Excitation spectrum, completeness and thermodynamics},
  journal       = {Int. J. Mod. Phys. A},
  volume        = {9},
  number        = {28},
  pages         = {4921--4948},
  year          = {1994},
  doi           = {10.1142/S0217751X94001977},
  eprint        = {hep-th/9310133},
  archivePrefix = {arXiv}
}

@article{gils2013anyonic,
  title = {Anyonic quantum spin chains: Spin-1 generalizations and topological stability},
  author = {Gils, C. and Ardonne, E. and Trebst, S. and Huse, D. A. and Ludwig, A. W. W. and Troyer, M. and Wang, Z.},
  journal = {Phys. Rev. B},
  volume = {87},
  issue = {23},
  pages = {235120},
  numpages = {33},
  year = {2013},
  month = {Jun},
  publisher = {American Physical Society},
  doi = {10.1103/PhysRevB.87.235120},
  url = {https://link.aps.org/doi/10.1103/PhysRevB.87.235120}
}

@article{Fradkin1980Parafermion,
  author  = {Fradkin, Eduardo and Kadanoff, Leo P.},
  title   = {Disorder variables and para-fermions in two-dimensional statistical mechanics},
  journal = {Nucl. Phys. B},
  volume  = {170},
  number  = {1},
  pages   = {1--15},
  year    = {1980},
  doi     = {10.1016/0550-3213(80)90472-1}
}

@ARTICLE{Zamolodchikov1985Parafermion,
       author = {{Zamolodchikov}, A.~B. and {Fateev}, V.~A.},
        title = "Nonlocal (parafermion) currents in two-dimensional conformal quantum field theory and self-dual critical points in {$Z_N$}-symmetric statistical systems",
      journal = {Sov. Phys. JETP},
         year = 1985,
        month = aug,
       volume = {62},
       number = {2},
        pages = {215},
       url = {https://www.jetp.ras.ru/cgi-bin/e/index/e/62/2/p215?a=list}
}

@article{Fateev1982IntegrableZN,
title = {Self-dual solutions of the star-triangle relations in {$Z_N$}-models},
journal = {Phys. Lett. A},
volume = {92},
number = {1},
pages = {37-39},
year = {1982},
issn = {0375-9601},
doi = {https://doi.org/10.1016/0375-9601(82)90736-8},
url = {https://www.sciencedirect.com/science/article/pii/0375960182907368},
author = {V.A. Fateev and A.B. Zamolodchikov}
}

@article{Alcaraz1986QuantumIntegrableZN,
title = {Conservation laws for {$Z(N)$} symmetric quantum spin models and their exact ground state energies},
journal = {Nucl. Phys. B},
volume = {275},
number = {3},
pages = {436-458},
year = {1986},
issn = {0550-3213},
doi = {https://doi.org/10.1016/0550-3213(86)90608-5},
url = {https://www.sciencedirect.com/science/article/pii/0550321386906085},
author = {Francisco C. Alcaraz and A. {Lima Santos}}
}

@article{Polackova2023CTMRG,
title = {Anisotropic deformation of the 6-state clock model: Tricritical-point classification},
journal = {Physica A},
volume = {624},
pages = {128907},
year = {2023},
issn = {0378-4371},
doi = {https://doi.org/10.1016/j.physa.2023.128907},
url = {https://www.sciencedirect.com/science/article/pii/S0378437123004624},
author = {Maria Polackova and Andrej Gendiar}
}

@misc{Rakala2021TriangularIsingZ6,
      title={Melting of three-sublattice order in triangular lattice {Ising} antiferromagnets: Power-law order, {$Z_6$} parafermionic multicriticality, and weakly first order transitions}, 
      author={G. Rakala and N. Desai and S. Shivam and K. Damle},
      year={2021},
      eprint={2109.03178},
      archivePrefix={arXiv},
      primaryClass={cond-mat.stat-mech},
      url={https://arxiv.org/abs/2109.03178}, 
}

@article{Jose1977PlanarModel,
  title = {Renormalization, vortices, and symmetry-breaking perturbations in the two-dimensional planar model},
  author = {Jos\'e, Jorge V. and Kadanoff, Leo P. and Kirkpatrick, Scott and Nelson, David R.},
  journal = {Phys. Rev. B},
  volume = {16},
  issue = {3},
  pages = {1217--1241},
  numpages = {0},
  year = {1977},
  month = {Aug},
  publisher = {American Physical Society},
  doi = {10.1103/PhysRevB.16.1217},
  url = {https://link.aps.org/doi/10.1103/PhysRevB.16.1217}
}

@article{Ortiz2012PClockDualities,
  author  = {Ortiz, G. and Cobanera, E. and Nussinov, Z.},
  title   = {Dualities and the phase diagram of the {$p$}-clock model},
  journal = {Nucl. Phys. B},
  volume  = {854},
  number  = {3},
  pages   = {780--814},
  year    = {2012},
  doi     = {10.1016/j.nuclphysb.2011.09.012}
}

@article{Sun2019ZpClockTransitions,
  title = {Phase transitions in the {$\mathbb{Z}_{p}$} and {U(1)} clock models},
  author = {Sun, G. and Vekua, T. and Cobanera, E. and Ortiz, G.},
  journal = {Phys. Rev. B},
  volume = {100},
  issue = {9},
  pages = {094428},
  numpages = {16},
  year = {2019},
  month = {Sep},
  publisher = {American Physical Society},
  doi = {10.1103/PhysRevB.100.094428},
  url = {https://link.aps.org/doi/10.1103/PhysRevB.100.094428}
}

@article{Challa1986SixStateClock,
  title = {Critical behavior of the six-state clock model in two dimensions},
  author = {Challa, Murty S. S. and Landau, D. P.},
  journal = {Phys. Rev. B},
  volume = {33},
  issue = {1},
  pages = {437--443},
  numpages = {0},
  year = {1986},
  month = {Jan},
  publisher = {American Physical Society},
  doi = {10.1103/PhysRevB.33.437},
  url = {https://link.aps.org/doi/10.1103/PhysRevB.33.437}
}

@article{Matsuo2006SixStateClock,
doi = {10.1088/0305-4470/39/12/006},
url = {https://doi.org/10.1088/0305-4470/39/12/006},
year = {2006},
month = {mar},
publisher = {},
volume = {39},
number = {12},
pages = {2953},
author = {Matsuo, Haruhiko and Nomura, Kiyohide},
title = {{Berezinskii–Kosterlitz–Thouless} transitions in the six-state clock model},
journal = {J. Phys. A}
}

@Article{tenpy,
	title={Efficient numerical simulations with Tensor Networks: {Tensor Network Python (TeNPy)}},
	author={Johannes Hauschild and Frank Pollmann},
	journal={SciPost Phys. Lect. Notes},
	volume={5},
	year={2018},
	publisher={SciPost},
	doi={10.21468/SciPostPhysLectNotes.5},
	url={https://scipost.org/10.21468/SciPostPhysLectNotes.5},
}

@misc{colorcet,
      title={Good Colour Maps: How to Design Them}, 
      author={Peter Kovesi},
      year={2015},
      eprint={1509.03700},
      archivePrefix={arXiv},
      primaryClass={cs.GR},
      url={https://arxiv.org/abs/1509.03700}, 
      note={Ready-made colour maps at \url{https://colorcet.com}}
}

@misc{pyfssa,
      title={autoScale.py - A program for automatic finite-size scaling analyses: A user's guide}, 
      author={O. Melchert},
      year={2009},
      eprint={0910.5403},
      archivePrefix={arXiv},
      primaryClass={physics.comp-ph},
      url={https://arxiv.org/abs/0910.5403}, 
      note={{\tt pyfssa} implementation by Andreas Sorge, updated version at \url{https://github.com/attila-i-szabo/pyfssa}}
}

\clearpage

\section{End matter}

\subsection{Symmetries and dualities of the clock model}

The clock, Potts, and generalised models~\mbox{(\ref{eq: clock},\,\ref{eq: Potts}--\ref{eq: combined H})} are all invariant under the Kramers--Wannier duality
\begin{align}
    X^\dagger_{i}X_{i+1} &\leftrightarrow \tilde Z_{i+1/2}, &
    Z_i & \leftrightarrow \tilde X^\dagger_{i-1/2}\tilde X_{i+1/2}, &
    J_q&\leftrightarrow h_q.
\end{align}
This implies that their phase diagrams are symmetric around $J_q=h_q$:
Symmetry-broken phases are dual to ones where the corresponding order parameter correlators decay exponentially, while the Luttinger liquid is self-dual. The former implies that in paramagnetic phases, correlators and  (on open chains) expectation values of the dual order parameter $\tilde X$,
\begin{subequations}
\label{eq: dual OP}
\begin{align}
    \langle\tilde X^q_{i-1/2}\tilde X^{-q}_{j+1/2}\rangle &= \bigg\langle\prod_{k=i}^j Z_k^q\bigg\rangle, 
    \label{eq: dual OP correlation}\\
    \langle\tilde X^q_{i+1/2}\rangle &= \bigg\langle\prod_{k=1}^i Z_k^q\bigg\rangle,
    \label{eq: dual OP expectation}
\end{align}
\end{subequations}
are finite.

The Hamiltonian~\eqref{eq: combined H} reduces to the \textit{antiferromagnetic} clock model at $\alpha=0$. However, the phase diagram of the latter is equivalent to the ferromagnetic clock model~\eqref{eq: clock} for any even $p$: The unitary transformation
\begin{align}
    Z_i &\mapsto -Z_i,&
    X_{2i}&\mapsto -X_{2i},&
    X_{2i+1}&\mapsto X_{2i+1}
\end{align}
maps the two Hamiltonians on one another, respecting the commutation relations~\eqref{eq: commutation relation}.

\subsection{Low-energy spectrum from DMRG}

\begin{figure}
    \centering
    \includegraphics{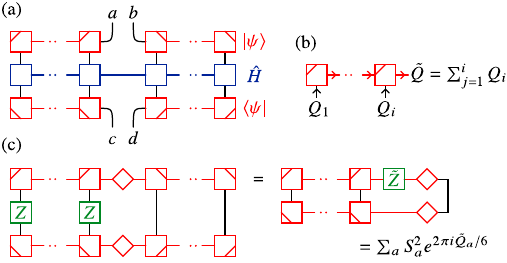}
    \caption{(a) Effective Hamiltonian $H^{cd}_{ab}$ for estimating the spectrum in \cref{fig: PBC}(b).
    (b) Illustration of the conserved $\z6$ charge on the virtual MPS indices.
    (c) Evaluating the string operator~\eqref{eq: dual OP expectation} in a $\z6$ symmetric MPS.
    Diagonal hatches in the MPS tensors indicate tensors in left or right canonical form. Red diamonds stand for diagonal matrices containing the Schmidt values in mixed canonical form.
    }
    \label{fig: MPS}
\end{figure}

To compute the spectrum in \cref{fig: PBC}(b), we deviated slightly from the approach of~\cite{Chepiga2017SpectrumDMRG}: Instead of changing the Lanczos diagonalisation within the DMRG loop, we diagonalised a ``zero-site'' effective Hamiltonian constructed from the converged MPS as illustrated in \cref{fig: MPS}(a). 
This corresponds to looking for the low-energy states in the subspace spanned by the leading Schmidt vectors, $|\psi_{ab}\rangle = |a_L\rangle\otimes|b_R\rangle$, of the ground state that are kept by DMRG.
Similar to the original approach, which differs in keeping the local Hilbert space of the orthogonality centre of the DMRG iteration intact, this truncation is not universally valid, but it works at conformal critical points.~\cite{Chepiga2017SpectrumDMRG}.

\subsection{Efficient string order parameters from DMRG}

We enforced the $\z6$ symmetry generated by $\hat Z$ in our simulations by assigning $\z6$ symmetry quantum numbers to the virtual legs of the MPS tensor as well and requiring that these quantum numbers satisfy a Gauss' law on all tensors~\cite{Singh2011U1SymmetricMPS}. 
It follows that the quantum numbers assigned to the virtual legs measure the total $\z6$ charge to the left of the bond in question, see \cref{fig: MPS}(b):
In particular, in an MPS in mixed canonical form, Schmidt values and vectors are labelled by their total $\z6$ charge $\tilde Q$.
Now, the dual order parameter expectation value~\eqref{eq: dual OP expectation} can be written as
\begin{equation}
    \langle\tilde X^q_{i+1/2}\rangle = \langle e^{2\pi i \tilde Q/6}\rangle = \sum_a S_a^2 e^{2\pi i \tilde Q_a/6},
\end{equation}
where the last sum runs over the Schmidt values $S_a$ of entanglement cut $i$, with associated charge $Q_a$.
That is, the dual order parameters plotted in \cref{fig: OBC}(b) can be computed efficiently from MPS ground state, using only the Schmidt values of a single entanglement cut per data point.

\end{document}


\title{Supplementary material to the paper\\``Parafermionic and decoupled multicritical points in a frustrated $\mathbb{Z}_6$ clock chain''}
\author{Andrea Kouta Dagnino}
\author{Attila Szabó}
\affiliation{Physik-Institut, Universität Zürich, Winterthurerstr.\ 190, 8057 Zürich, Switzerland}
\date{\today}

\maketitle

\section{Details of the DMRG simulations}

\begin{table}[b]
    \centering
    \setlength{\tabcolsep}{1em}
    \begin{tabular}{ccccccc}\hline\hline
        $J/h$ range & \#\,$J/h$ points & $\alpha$ range & \#\,$\alpha$ points & system size & bond dimension & BC\\\hline
        $0.5-2$ & 31 (log) & $0-1$ & 50 & 240 & 384 & open\\
        $0.36-0.44$ & 17 & $0-0.5$ & 17 & 240 & 384 & open\\
        $0.9-1$ & 20 & $0.355-0.37$ & 40 & 240 & 384 & open\\\hline
        && \multicolumn{2}{l}{$0.350-0.358\ \text{@}\ \Delta\alpha=0.0005,$} & $192, 240, 300$ & 384 & open\\
        \multicolumn{2}{c}{$J=h$} & \multicolumn{2}{c}{$0.358-0.362\ \text{@}\ \Delta\alpha=0.0002,$\raisebox{0pt}[0pt][0pt]{$\left.\rule{0pt}{2.25em}\right\}$}} & 384 & 480 & open\\
        && \multicolumn{2}{l}{$0.362-0.366\ \text{@}\ \Delta\alpha=0.0005$} & $480, 600$ & 600 & open\\\hline
        \multicolumn{2}{c}{$J=h$} & \multicolumn{2}{c}{$0.3606$} & $48,60$ & 1920 & periodic \\\hline
    \end{tabular}
    \caption{Summary of the DMRG scans discussed in this work.}
    \label{tab: dmrg details}
\end{table}

Parameters of the DMRG simulations reported in this paper are summarised in \cref{tab: dmrg details}. For the simulations near the parafermionic multicritical point, we adjusted the bond dimension with system size so as to keep the truncation error of two-site DMRG just below $10^{-8}$.

\section{Comprehensive plots of direct and string order parameters}

\begin{figure}[!b]
    \centering
    \includegraphics{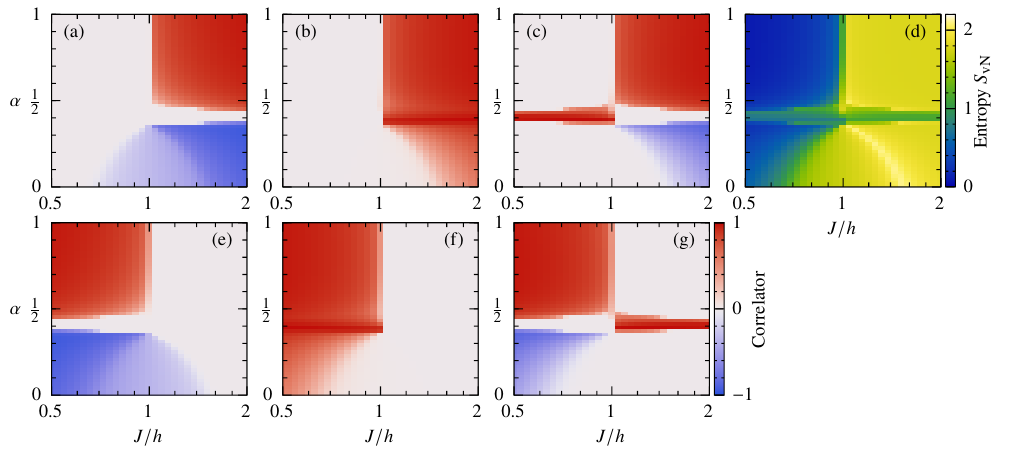}
    \caption{Correlators $\langle X_{60}X^\dagger_{181}\rangle$ (a), $\langle X^2_{60}X^{-2}_{181}\rangle$ (b), $\langle X^3_{60}X^3_{181}\rangle$ (c), half-chain entanglement entropy (d), and dual string correlators $\langle \prod_{i=60}^{180} Z_i\rangle$ (e), $\langle \prod_{i=60}^{180} Z_i^2\rangle$ (f), $\langle \prod_{i=60}^{180} Z_i^3\rangle$ (g) of the generalised clock model on a 240-site chain for $1/2\le J/h\le 2$, $0\le\alpha\le 1$.}
    \label{fig: overview}
\end{figure}

\begin{figure}
    \centering
    \includegraphics{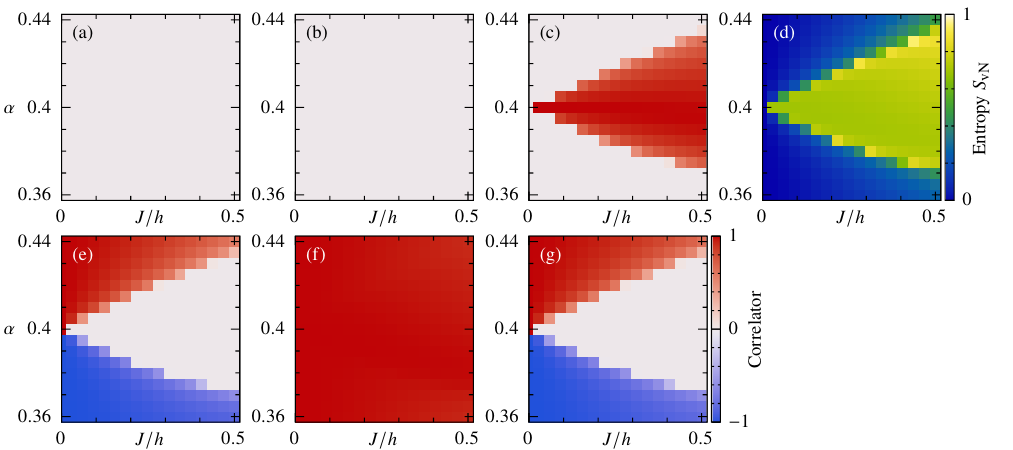}
    \caption{Correlators $\langle X_{60}X^\dagger_{181}\rangle$ (a), $\langle X^2_{60}X^{-2}_{181}\rangle$ (b), $\langle X^3_{60}X^3_{181}\rangle$ (c), half-chain entanglement entropy (d), and dual string correlators $\langle \prod_{i=60}^{180} Z_i\rangle$ (e), $\langle \prod_{i=60}^{180} Z_i^2\rangle$ (f), $\langle \prod_{i=60}^{180} Z_i^3\rangle$ (g) of the generalised clock model on a 240-site chain at low $J/h$ near the spin-flop transition at $J=0,\alpha=2/5$.}
    \label{fig: lobe}
\end{figure}

\begin{figure}
    \centering
    \includegraphics{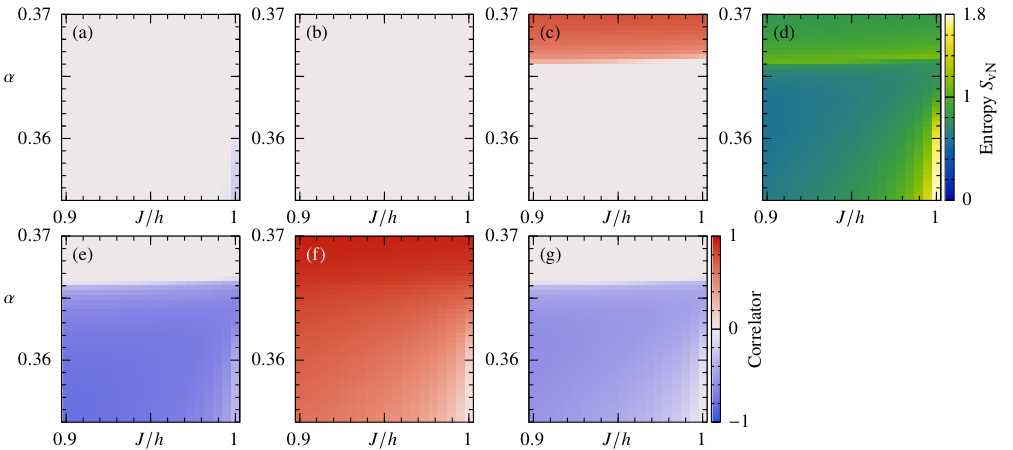}
    \caption{Correlators $\langle X_{60}X^\dagger_{181}\rangle$ (a), $\langle X^2_{60}X^{-2}_{181}\rangle$ (b), $\langle X^3_{60}X^3_{181}\rangle$ (c), half-chain entanglement entropy (d), and dual string correlators $\langle \prod_{i=60}^{180} Z_i\rangle$ (e), $\langle \prod_{i=60}^{180} Z_i^2\rangle$ (f), $\langle \prod_{i=60}^{180} Z_i^3\rangle$ (g) of the generalised clock model on a 240-site chain in the confluence region $J\approx h, \alpha\approx 0.36$.}
    \label{fig: zoom}
\end{figure}

\cref{fig: overview,fig: lobe,fig: zoom} show all standard and string correlation functions between quarter-chain points obtained from the DMRG scans shown in the main text. In particular, \cref{fig: overview} demonstrates the interchange between standard and string order parameters under Kramers--Wannier duality, which reverses the $J/h$ axis on the log scale used in the figure.

